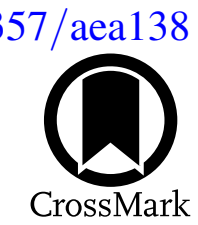

# Prompt and Afterglow Constraints on the Central Engine of GRB 240825A

Daban Mohammed Saeed[1,2], Rahim Moradi[3,1,4], Chen-Wei Wang[1,2], Brwa Shukur Ismael[1,2], and Marco Muccino[5,6,3]
[1] State Key Laboratory of Particle Astrophysics, Institute of High Energy Physics, Chinese Academy of Sciences, Beijing 100049, People's Republic of China; cwwang@ihep.ac.cn, rahim.moradi@ihep.ac.cn
[2] University of Chinese Academy of Sciences, Chinese Academy of Sciences, Beijing 100049, People's Republic of China
[3] ICRANet, Piazza della Repubblica 10, 65122 Pescara, Italy
[4] ICRA, Dip. di Fisica, Sapienza Università di Roma, Piazzale Aldo Moro 5, I-00185 Roma, Italy
[5] Università di Camerino, Via Madonna delle Carceri, Camerino 62032, Italy
[6] Al-Farabi Kazakh National University, Al-Farabi av. 71, 050040 Almaty, Kazakhstan

## Abstract

GRB 240825A shows unprecedented temporal and spectral features that constrain the properties of its inner engine: (1) a prompt emission three-component spectrum consisting of a Band function continuum, a quasi-thermal bump, and a hard MeV tail; and (2) an X-ray afterglow fitted by a power-law decay with index $1.29 \pm 0.02$; and (3) a $6.37 \pm 0.05$ Hz quasi-periodic oscillation (QPO) identified in the 100–300 keV band during the 2.07–3.25 s time interval, coinciding with the photospheric radiation phase. We compare two central-engine candidates: (a) a newborn millisecond magnetar undergoing multipolar spin-down, free precession, and global magnetoelastic oscillations of its interior; and (b) a Kerr black hole powering a Blandford–Znajek jet subject to Lense–Thirring disk precession. In principle, both channels can reproduce the observed QPO frequency but impose different constraints on the energetics, temporal evolution, and the origin of the oscillation. We first demonstrate that the X-ray afterglow is quantitatively reproduced by a hexapolar magnetar spin-down model with an initial spin period of $P_0 \simeq 1.38$ ms and a surface field of $B_{\rm hexa} \simeq 2.04 \times 10^{16}$ G. The decay index measures a braking index rather than the field multipolarity, does not exclude a fallback-regulated flow, and leaves $P_0$ and $B_{\rm hexa}$ uncertain by a factor of a few. In this scenario, magnetic or magnetoinertial dynamics of the star interior provide a plausible explanation of the QPO without invoking extreme stellar deformations. Although a black hole engine cannot be firmly excluded, the combined prompt and afterglow observational properties of GRB 240825A favor a long-lived magnetar central engine.

*Unified Astronomy Thesaurus concepts:* Gamma-ray bursts (629); Magnetars (992); Kerr black holes (886)

## 1. Introduction

Gamma-ray bursts (GRBs) are powered by compact central engines whose nature, whether a rapidly spinning magnetar or an accreting Kerr black hole, remains an open problem in high-energy astrophysics (T. Piran 1999; B. D. Metzger 2010; P. Kumar & B. Zhang 2015; B. Zhang 2018).

GRB events exhibiting rich observational features, such as displaying multiple well-separated spectral components, long-lived central-engine activities, and quasi-periodic oscillations (QPOs), are particularly valuable for diagnosing the nature of the central engine. In this context, time-resolved spectral analysis can disentangle specific, distinctive components and illustrate the spectral evolution of the prompt emission.

While the majority of GRB prompt emission spectra are described by a smoothly broken power-law function, known as the Band function (D. Band et al. 1993), some spectra can only be reproduced with additional components, such as a thermal component typically associated with the photospheric region of the jet (F. Ryde & A. Pe'er 2009; M. Ajello et al. 2019), and an extra nonthermal MeV–GeV tail attributed to synchrotron or inverse Compton radiation (F. Ryde & A. Pe'er 2009; M. Ajello et al. 2019). Studying the evolution of these spectral components discloses information about jet magnetization, baryon loading, radiative efficiency, and the location of dissipation (see e.g., C.-W. Wang et al. 2025, and references therein).

On longer timescales, a large fraction of GRBs (>50% in long GRB and between 18% and 37% in short GRBs) following an initial steep decay exhibit an early afterglow shallow decay known as the plateau phase (L. Guglielmi et al. 2024). Such plateau features are widely interpreted as evidence for extended central-engine activity. Magnetar spin-down (S. Dall'Osso et al. 2011; A. Rowlinson et al. 2013; S. X. Yi et al. 2014; M. G. Dainotti et al. 2017; Y. Wang et al. 2024; E. S. Yorgancioglu et al. 2025) and black hole spin-down through a magnetized disk (P. Kumar & B. Zhang 2015; W.-H. Lei et al. 2017; A. Ł. Lenart et al. 2025) both emerge as plausible central-engine candidates, although they predict different energetics, energy-injection histories, and possible multimessenger signatures.

Additionally, the detection of QPOs might provide a more direct probe of the central engine, since their characteristics naturally reflect physical timescales of the compact object or its surrounding environment. In magnetar scenarios, proposed origins include free-precession, magnetoelastic, or Alfvén modes (Y. Levin 2006; H. Sotani et al. 2008); meanwhile, proposed origins in black hole models include Lense–Thirring (LT) disk precession (J. Lense & H. Thirring 1918; P. C. Fragile et al. 2007; N. Stone & A. Loeb 2012). Although QPO candidates have been reported in a limited number of GRBs (C. Chirenti et al. 2023; R.-C. Chen et al. 2025), clear detections associated with specific emission components remain rare.

GRB 240825A is a rich observational source in this context, providing a fascinating window into the nature of the inner engine. Its prompt emission contains three well-resolved components (a) a quasi-thermal photosphere, (b) a Band-like nonthermal spectrum, and (c) a hard MeV tail, which were observed simultaneously over a broad energy range (C.-W. Wang et al. 2025). The burst also shows a steep early afterglow with a subsequent shallower decay (C. Wu et al. 2025), and a $6.37 \pm 0.05$ Hz QPO detected specifically in the quasi-thermal photospheric component during the 2.07–3.25 s interval (G.-Y. Li et al. 2025).

In this work, we exploit these observations to perform a detailed comparison between two leading central-engine candidates: (1) a newborn magnetar with spin-down and possible precession or magnetoelastic oscillation modes; and (2) a Kerr black hole powering a Blandford–Znajek (BZ) jet subject to LT precession. We evaluate and compare how effectively each model can simultaneously comprise the aforementioned broadband features of GRB 240825A, while stating the limitations and physical requirements of the framework.

## 2. Spectral Analysis and Summary of Existing Observations and Interpretations

GRB 240825A is a bright GRB with an intermediate duration detected by many gamma-ray monitors, including GECAM-B (C.-W. Wang et al. 2024), Insight-HXMT, Fermi (V. Sharma & C. Meegan 2024), and Swift-BAT (R. Gupta et al. 2024) at 15:53:00.850 UT on 2024 August 25 (denoted as $T_0$). Rapid and continuous multiwavelength follow-up observations were conducted by Swift-XRT/UVOT (R. Gupta et al. 2024), SVOM-VT (Y. L. Qiu et al. 2024), as well as a series of ground-based telescopes. Host-galaxy observations suggest that the burst occurred in a massive, dusty, star-forming system, with no bright supernova detected down to deep limits (R. Gupta et al. 2025). The redshift is measured as $z = 0.659$ by the Very Large Telescope (VLT) (A. Martin-Carrillo et al. 2024) and Telescopio Nazionale Galileo (TNG) (A. Melandri et al. 2024).

Using joint analyses of Fermi-GBM/LAT, GECAM, Insight-HXMT, and rapid optical–near-infrared (NIR) follow-up, several groups have already analyzed GRB 240825A, capturing a coherent observational picture that shapes our physical interpretation. The time-resolved spectral analysis by C.-W. Wang et al. (2025) shows that the prompt emission requires three distinct radiative components. The earliest pulses are characterized by a subdominant quasi-thermal photospheric bump with $kT \sim 30$ keV, and vanish before the onset of GeV emission. Meanwhile, a Band-like nonthermal component with $E_{\rm peak} \sim 400$ keV governs the main luminosity peak. During the second major pulse, an additional high-energy component appears and exhibits an intrinsic spectral break at $\sim$40–50 MeV. This multicomponent structure produces the characteristic two-hump spectral shape reported in earlier studies (H.-M. Zhang et al. 2025) and cannot be reproduced by any single Band function. The relative evolution of these three components indicates changes in dissipation radii, Lorentz factor, and baryon loading, naturally motivating a multizone jet framework in which the photosphere, internal dissipation region, and outer MeV–GeV zone are physically distinct.

While the three-component prompt spectral evolution is naturally accommodated within hyperaccreting black hole jet models, it does not uniquely determine the compact-object identity (C.-W. Wang et al. 2025). A magnetar central engine therefore, remains viable. In this picture, the early quasi-thermal component can be associated with a baryon-loaded protomagnetar outflow, whereas the subsequent decrease of baryon loading and increase of magnetization can enable magnetic dissipation and prolonged energy injection into the afterglow. This provides a physically continuous framework linking the early photospheric emission, the later nonthermal activity, and the magnetar-driven afterglow interpretation (B. D. Metzger et al. 2011; A. Rowlinson et al. 2013; P. Beniamini et al. 2017).

Moreover, a key temporal feature reported by G.-Y. Li et al. (2025) is the detection of a high-precision QPO at $6.37 \pm 0.05$ Hz with quality factor $Q \sim 5$ and fractional rms of a few to $\sim$10%, hinting at a quasi-periodic oscillating jet. Importantly, this modulation is detected only in the 100–300 keV band, making it the first ever detection of QPOs confined to the thermal component, and not to the nonthermal components. This strongly suggests that the oscillation originates at or below the photospheric radius and is therefore directly linked to the conditions established by the central engine.

R. Gupta et al. (2025) report a softer extended-emission tail in the Swift-BAT light curve that lasts up to $\sim T_0 + 300$ s. This is usually interpreted as continued activity of the central (B. D. Metzger et al. 2008; B. P. Gompertz et al. 2013). At the same time, the X-ray and optical afterglows cannot be described together with a single standard synchrotron closure relation, which points to a more complex afterglow structure.

A more detailed interpretation is given by C. Wu et al. (2025), who model the early emission using a forward–reverse shock framework. In their picture, the early optical–NIR emission is dominated by the reverse shock, and later the forward shock takes over, aligned with the standard reverse-to-forward transition seen in GRB afterglows (B. Zhang et al. 2003). The inferred magnetic field ratio between the reverse and forward shock regions, is $R_B \sim 300$, suggesting that the reverse shock region is much more magnetized than the forward shock region. The simplest interpretation is that the ejecta already carries a strong magnetic field from the central engine, possibly a magnetar, rather than generating it only through shock processes.

Taken together, these observations establish three key empirical features of GRB 240825A: (1) a distinctly resolved three-component prompt spectrum, (2) a narrow photospheric QPO during the early emission, and (3) a substantial shallowing of the x-ray and optical afterglow, with evidence for a reverse shock. These characteristics make GRB 240825A a proper probe for investigating central-engine physics, offering a unique opportunity to link prompt spectral structure, early-time variability, and long-lived energy injection within a cohesive physical framework.

In the remainder of this section, we present our own spectral modeling and energetic analysis, using a uniform treatment of the Fermi Gamma-Ray Space Telescope data to derive the luminosity and isotropic energy.

The Fermi telescope comprises two instruments: the Gamma-Ray Burst Monitor (GBM; C. Meegan et al. 2009) and the Large Area Telescope (LAT; W. B. Atwood et al. 2009). Fermi/GBM is equipped with 12 sodium iodide (NaI) scintillation detectors and two bismuth germanate (BGO)

**Table 1**
Spectral Fit Results of Time-integrated Spectra for the Band, CPL, PL, and Band+BB Models

| Model | $\alpha$ | $\beta$ | $E_p$ (keV) | Norm | $kT$ (keV) | $S_{\rm bolo}$ ($\times10^{-4}$ erg cm$^2$) | $E_{\rm iso}$ ($\times10^{53}$ erg) | BIC | AIC |
|---|---|---|---|---|---|---|---|---|---|
| Band | $-0.82^{+0.01}_{-0.01}$ | $-2.24^{+0.02}_{-0.02}$ | $438^{+15}_{-15}$ | 0.25 | ... | $1.46^{+0.02}_{-0.02}$ | $2.02^{+0.03}_{-0.03}$ | 490 | 471 |
| CPL | $-0.87^{+0.01}_{-0.01}$ | ... | $566^{+24}_{-24}$ | $17.47^{+0.48}_{-0.48}$ | ... | $1.18^{+0.04}_{-0.04}$ | $1.42^{+0.05}_{-0.05}$ | 1199 | 1214 |
| PL | $-1.33^{+0.00}_{-0.00}$ | ... | ... | $91.61^{+1.55}_{-1.55}$ | ... | $2.86^{+0.04}_{-0.04}$ | $3.43^{+0.05}_{-0.05}$ | 10360 | 10371 |
| Band+BB | $-0.75^{+0.03}_{-0.03}$ | $-2.23^{+0.02}_{-0.02}$ | $415^{+23}_{-23}$ | $0.26^{+0.01}_{-0.01}/2.58^{+1.20}_{-1.20}$ | $8.23^{+0.88}_{-0.88}$ | $1.39^{+0.04}_{-0.04}$ | $2.02^{+0.05}_{-0.05}$ | 492 | 466 |

**Note.** (1) The Band function is a smoothly joined broken power law (D. Band et al. 1993; B. Zhang 2018), with low- and high-energy photon indices $\alpha$ and $\beta$, respectively. The peak energy $E_{\rm p}$ is related to the break energy via $E_{\rm p} = (2 + \alpha)E_0$. (2) The CPL model represents the low-energy branch of the Band function, with $E_{\rm p} = (2 + \alpha)E_c$ Y. Kaneko et al. (2008), J. F. Steiner et al. (2009). (3) The PL model is described by a single photon index $\alpha$. (4) For the Band+BB model, the reported normalization values correspond to the Band and blackbody components, respectively.

detectors sensitive to photons in the energy range 8 keV–40 MeV.

In our analysis, the GBM data for GRB 240825A were extracted from the Fermi Science Support Center (FSSC).[7] We utilized the Fermi GBM Data Tools (GDT; A. Goldstein et al. 2024), an analysis toolkit that facilitates the reduction of Fermi-GBM data. Two NaI detectors (n6 and n7) with incident angles less than 60° were selected, along with one BGO detector (b1) with an incident angle less than 80°. For both detector types, background time intervals were manually selected before and after GRB prompt emission phase and fitted with a first-degree polynomial, enabling the extraction of the spectral range. The $T_{90}$ duration, the time taken to accumulate 90% of the burst fluence starting at the 5% fluence level, is around 4 s for GRB 240825A, starting from 1.152 s to 5.12 s. Subsequently, to calculate the isotropic-equivalent energy, we adopted a redshift of $z = 0.659$ for GRB 240825A, measured by VLT/X-shooter (A. Martin-Carrillo et al. 2024). To perform the spectral analysis over the full $T_{90}$ interval, we utilized XSPEC (K. A. Arnaud 1996), a spectral fitting package widely used due to its large number of theoretical models. Several spectral models, such as Band (D. Band et al. 1993), cutoff power law (CPL; Y. Kaneko et al. 2008; J. F. Steiner et al. 2009), power law (PL), and blackbody (G. Ghirlanda et al. 2003), were used as a single or multicomponent spectral model; see Appendix C. The best spectral fit and the parameter values of each model are listed in Table 1.

To judge which model yields the best results, we adopt the Akaike information criterion (AIC), defined by H. Akaike (1974) as AIC $= 2\ln L(\theta) + 2k$, and the Bayesian information criterion (BIC), derived by G. Schwarz (1978) as BIC $= 2\ln L(\theta) + k\ln(n)$, where $L(\theta)$ is the maximum likelihood of the model, $k$ is the number of free parameters, and $n$ is the number of data points. The usual procedure is to apply single-component spectral models (Band, blackbody (*BB*), *PL*, *CPL*) and select the model that minimizes AIC and BIC. Both criteria penalize model complexity, unless there is a significant improvement in the fit (R. E. Kass & A. E. Raftery 1995). In cases where $\Delta$AIC or $\Delta$BIC $\geqslant 6$, there is strong evidence indicating the improvement of the model. For more illustrative examples, see also Y.-Z. Wang et al. (2017), V. Chand et al. (2018), and T. Aldowma et al. (2026).

In our case, we employed both AIC and BIC to compare non-nested models, selecting the model with the lowest information-criterion value as our baseline model. Then, we construct nested models by supplementing additional components to the baseline model and compute $\Delta$AIC and $\Delta$BIC relative to the baseline to assess whether the added complexity is statistically justified (K. P. Burnham & D. R. Anderson 2004). In the end, we adopt the model (nested or non-nested) with the lowest information-criterion value to calculate the bolometric fluence ($S_{\rm bolo}$), defined as the total energy per unit area integrated over the $T_{90}$ interval, in the 8 keV–40 MeV band. We also compute the isotropic-equivalent energy ($E_{\rm iso}$), defined as the total energy assuming isotropic emission (see Appendix A).

In Figure 1, the $\nu F_\nu$ spectrum of GRB 240825A, obtained from two NaI detectors (n6 and n7) and one BGO detector (b1), is best fitted by the Band model over the energy range 10 keV–30 MeV, with the spectral peak well constrained at $E_{\rm peak} \simeq 438 \pm 15$ keV, a low-energy photon index of $\alpha = -0.82$, and a high-energy photon index of $\beta = -2.24$. The dominance of a single Band component indicates a nonthermal spectrum, consistent with synchrotron radiation scenarios. This time-integrated, GBM-only result does not contradict the three-component prompt-emission evolution reported by C.-W. Wang et al. (2025), since subdominant components are diluted by temporal integration and the delayed GeV component lies outside the GBM bandpass. Accordingly, we use the Band fit here solely to derive the burst energetics.

## 3. Magnetar Interpretation

### *3.1. Magnetar Spin-down and X-ray Afterglow of GRB 240825A*

Magnetars can exhibit dipolar and multipolar magnetic field components. For a dipole, the magnetic field strength can exceed or fall below the quantum-critical field $B_Q \approx 4.4 \times 10^{13}$ G, above which the polarization of the QED vacuum modifies photon propagation and pair processes (J. Schwinger 1951), whereas near-surface multipolar components can be even stronger, often reaching supercritical values of a few $\times10^{15}$ G. (P. Goldreich & A. Reisenegger 1992; U. Geppert et al. 1999; C. Thompson et al. 2002; N. Rea et al. 2012; K. N. Gourgouliatos et al. 2016; E. Göğüş et al. 2020). Despite the evidence for dominance of multipolar components in the early phases ($t \lesssim 10^6$ s), and their initially stronger magnetic fields, the dipole component will eventually ($t \gtrsim 10^8$ s) overtake the spin-down luminosity (see e.g., Y. Wang et al. 2024, and references therein). This is because the radiation losses of multipolar components are more sensitive to the spin rate of the magnetar, scaling as of $L_l \propto B_l^2\Omega^{2l+2}R^{2l+4}$, as established by the general formula for multipolar spin-down luminosity (J. Petri 2015; Y. Wang et al. 2024).

[7] https://fermi.gsfc.nasa.gov/ssc/data/access/gbm/

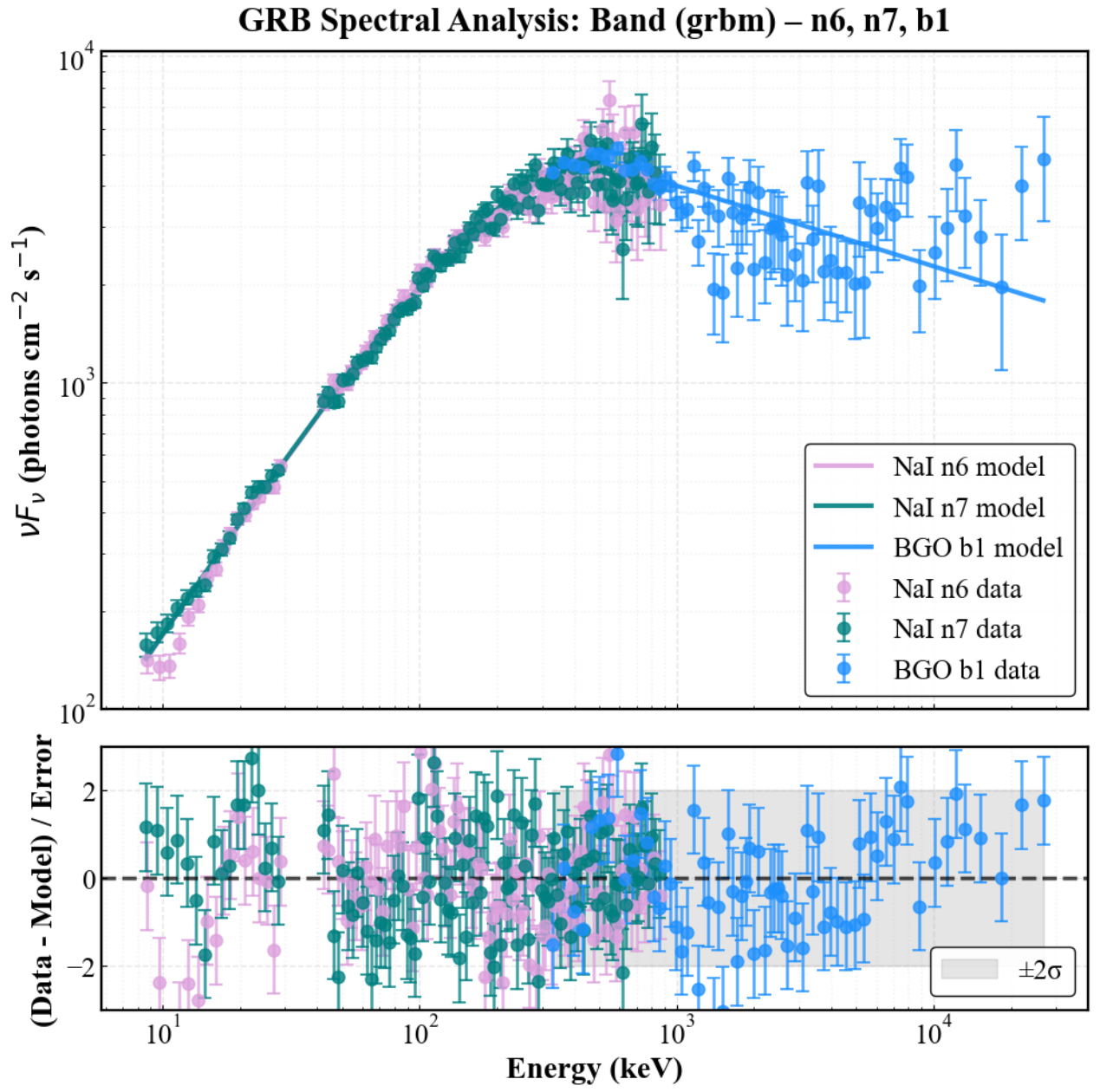

**Figure 1.** The $\nu F_\nu$ spectrum of GRB 240825A obtained with Fermi/GBM. The data from three detectors, n6, n7, and b1, are best fitted by a single spectral component, the Band function, with $E_{\rm peak} \sim 438$ keV.

$$L_l(t) = \frac{C_l\ \Omega^{2l+2} B_l^2 R^{2l+4} \Theta_l^2}{c^{\ 2l+1}}, \tag{1}$$

where $C_l$ is a constant term determined by the vacuum solution (2/3 for dipole, 32/135 for quadrupole, $2/4725\pi$ for hexapole, and $4/297675\pi$ for octapole), $\Omega$ is the angular velocity of the magnetar, $B_l$ is the magnetic field, $R$ is the radius, typically around $10^6$ cm, while $\Theta$ is the geometric inclination factor. The magnetar can consist of several orders of vector spherical harmonics, in which $l = 1$ for dipole, $l = 2$ for quadrupole, $l = 3$ for hexapole, and $l = 4$ for octupolar (R. G. Barrera et al. 1985; J. D. Jackson 1998).

Thus, the total multipolar spin-down luminosity is given as

$$L_{\rm tot}(t) = \sum_{l=1}^{\infty} L_l(t). \tag{2}$$

The two expressions that follow are obtained by truncating this sum at a single order. They therefore hold only while one multipole order $l$ dominates the spin-down over the interval of interest, an approximation we adopt following Y. Wang et al. (2024). If two orders contribute comparably, the braking is no longer a single power law and $\Omega(t)$ must be integrated from the full sum. The rate of change of rotational energy gives the time evolution of angular velocity

$$\Omega_l(t) = \Omega_{l,0}\left(1 + \frac{t}{\tau_l}\right)^{-1/(2l)}. \tag{3}$$

Substituting this result into Equation (1) returns

$$L_l(t) = L_{l,0}\left(1 + \frac{t}{\tau_l}\right)^{-(1+\frac{1}{l})}. \tag{4}$$

Here, $\tau_l$ is the characteristic spin-down timescale, which corresponds to the duration of the plateau, and $L_{l,0}$ is the initial spin-down luminosity

$$\tau_l = \frac{Ic^{2l+1}}{2l\ C_l \Omega_{l,0}^{2l} B_l^2 R^{2l+4} \Theta_l^2}, \tag{5}$$

$$L_{l,0} = \frac{I\Omega_{l,0}^2}{2l\ \tau_l}. \tag{6}$$

Both expressions follow from integrating $I\Omega\dot{\Omega} = -L_l$ with Equation (1).

The X-ray luminosity evolution of GRB 240825A in the 0.3–10 keV energy range, as observed by Swift (P. A. Evans et al. 2007; P. A. Evans et al. 2009), is shown in Figure 2. Throughout the afterglow modeling we fix the stellar parameters to their canonical values, $M = 1.4\,M_\odot$, $R = 1 \times 10^6$ cm, and $I = 10^{45}$ g cm$^2$, so that $B_l$ and $\Omega_{l,0}$ are the only free parameters of the spin-down law. The data are well fitted by a spin-down model dominated by a higher-order multipolar component rather than the typical dipole model. We find that the hexapole component provides the best description of the decay. The fitting was performed using the LMFIT library in Python, a tool for nonlinear least-squares minimization and curve fitting (M. Newville et al. 2014). The results confirm that supplementing additional components, such as a dipole, does not significantly improve the fit statistics. The lower panels show the residuals and the posterior distributions of the model parameters obtained from Markov Chain Monte Carlo (MCMC) analysis. We infer an initial spin period of ∼1.38 ms and a magnetic field strength of $2.04 \times 10^{16}$ G.

The quantity these numbers rest on is the decay index, and its relation to $l$ is worth making explicit. The local logarithmic slope of Equation (4) is

$$-\frac{d\ln L_l}{d\ln t} = \left(1 + \frac{1}{l}\right)\frac{t/\tau_l}{1 + t/\tau_l}, \tag{7}$$

which grows monotonically from zero for $t \ll \tau_l$ and reaches the asymptotic value $1 + 1/l$ for $t \gg \tau_l$. An observed decay index is therefore a lower bound on the asymptotic index, and the dominant order obeys

$$l \leqslant \frac{1}{\alpha_X - 1} = 3.4 \pm 0.2, \tag{8}$$

with equality in the asymptotic regime. For GRB 240825A, where $\alpha_X = 1.29 \pm 0.02$, this excludes an octupolar wind outright: its asymptotic index, $5/4 = 1.25$, is shallower than the observed decay, so it can never decay as steeply as observed. The remaining orders are separated by the curvature they predict. A given order reproduces a local slope of 1.29 at $t/\tau_l = \alpha_X/(1 + 1/l - \alpha_X)$, which is 1.8 for the dipole, 6.1 for the quadrupole, and 29.8 for the hexapole, and the slope must then steepen by a further 55%, 16%, and 3%, respectively, to reach its asymptote. The index measured for GRB 240825A is constant to $\pm 0.02$ across the fitted range and only the hexapole is close enough to its asymptote to remain so; this is also why the fit does not improve when a dipole component is added. Equation (8) nevertheless constrains a braking index: it fixes how steeply the engine power declines, and identifying it with a particular vector spherical harmonic follows only once the vacuum multipolar spin-down law of Equations (1)–(4) is assumed.

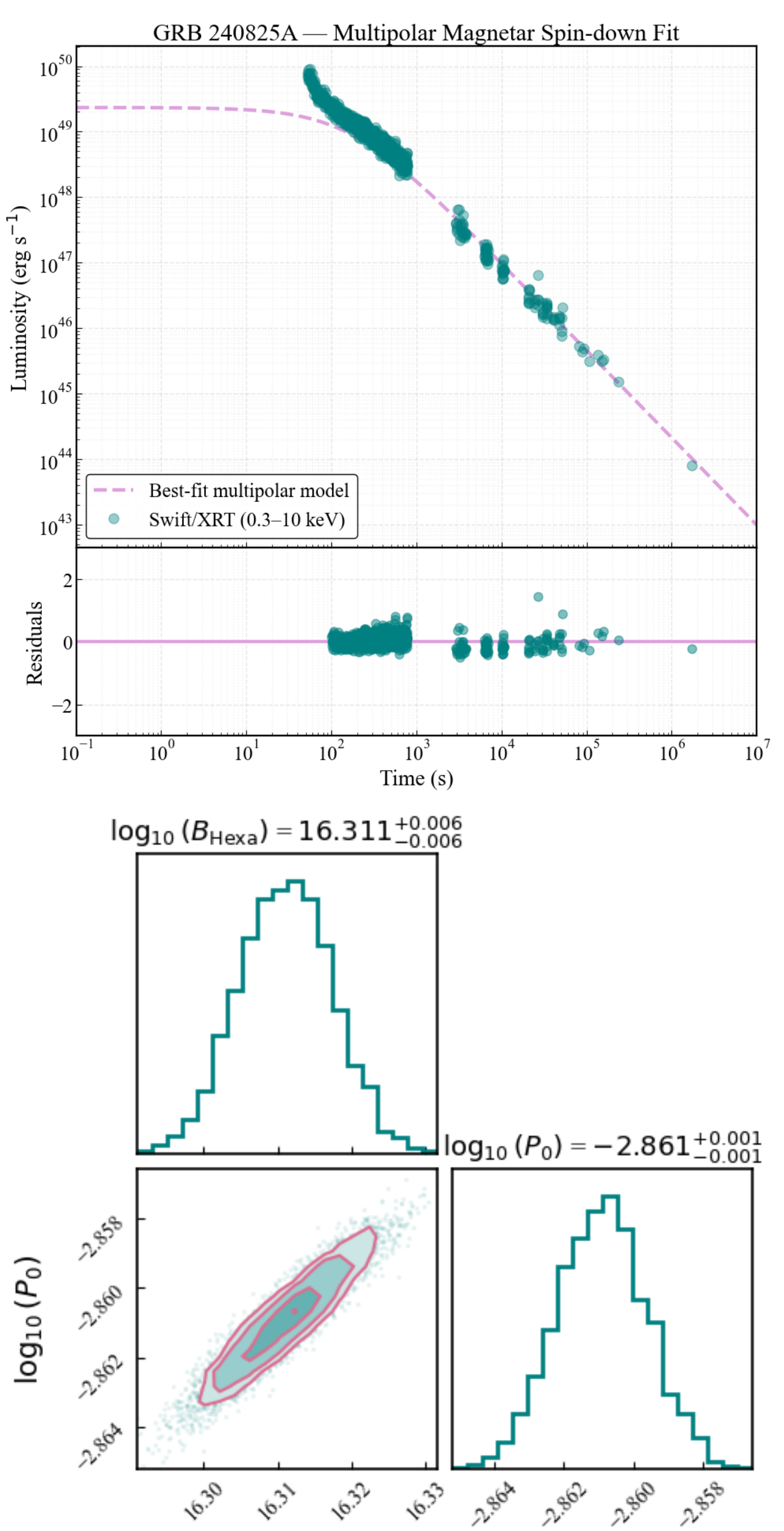

**Figure 2.** The spin-down luminosity inferred from Swift/XRT data (0.3–10 keV). The upper panel shows the luminosity evolution well described by a hexapole component rather than a dipole model. The lower panel shows the MCMC analysis inferring an initial spin period of $P_0 \simeq 1.38$ ms and a hexapolar magnetic field strength of $B_{\rm hexa} \simeq 2.04 \times 10^{16}$ G.

Two further assumptions enter the normalization. The fit above treats the 0.3–10 keV luminosity as isotropic and as tracking the spin-down power with unit efficiency. Neither holds in general. Following E. S. Yorgancioglu et al. (2025), the engine power implied by an observed isotropic-equivalent X-ray luminosity $L_{\rm sd}$ is

$$L_{X,\rm true} = \frac{\theta_{\rm j}^2}{2}\frac{L_{\rm sd}}{\eta_X}k, \tag{9}$$

where $\theta_{\rm j}$ is the jet half-opening angle, $\eta_X$ the X-ray radiative efficiency, and $k$ the bolometric correction. It is convenient to collect these into a single rescaling factor

$$f \equiv \frac{\theta_{\rm j}^2}{2}\frac{k}{\eta_X}, \tag{10}$$

so that our fit corresponds to $f = 1$. The important point for the present discussion is that $f$ is time independent to leading order, and therefore rescales $L_{l,0}$ without touching the shape of the light curve. The decay index, and with it $l_{\rm eff}$, is unaffected. What does change are the two derived parameters: holding the measured break time $\tau_l$ fixed, Equations (5) and (6) give $\Omega_{l,0} \propto f^{1/2}$ and $B_l \propto f^{-l/2}$, that is

$$P_0 \simeq 1.38 f^{-1/2}\ \text{ms}, \qquad B_{\rm hexa} \simeq 2.04 \times 10^{16} f^{-3/2}\ \text{G}. \tag{11}$$

The two quantities therefore move in opposite directions along this one-parameter family, which is relevant because an initial period close to mass-shedding and a field above $10^{16}$ G are both at the edge of what newborn magnetar models accommodate. Taking $f = 1.9$ gives $P_0 \simeq 1.0$ ms with a much more modest $B_{\rm hexa} \simeq 8 \times 10^{15}$ G, while $f = 0.5$ relaxes the period to $P_0 \simeq 1.95$ ms at the cost of $B_{\rm hexa} \simeq 6 \times 10^{16}$ G. The two corrections entering $f$ act in opposite senses. Beaming lowers the engine power required to explain a given observed flux and drives $f$ below unity, while a radiative efficiency $\eta_X < 1$ and a bolometric correction $k > 1$ raise it, so that they largely cancel in Equation (10). Y. Wang et al. (2024) adopt the same normalization on these grounds and estimate the residual uncertainty at a factor of about three. For a jet half-opening angle of a few degrees, Equation (10) then returns $f \simeq 1$ for $\eta_X \simeq 0.3\%$–$1\%$ if the fitted luminosity is the 0.3–10 keV band luminosity and $k \simeq 5$ is taken from Y. Wang et al. (2024), or for $\eta_X \simeq 0.06\%$–$0.2\%$ if the bolometric luminosity is used directly and $k = 1$. Both lie inside the 0.05%–10% range explored by R. Moradi et al. (2025) and below the ceiling found by D. Xiao & Z.-G. Dai (2019). Propagating the factor-of-three spread through Equation (11), and imposing $P_0 \gtrsim 1$ ms, gives $P_0 \simeq 1.0$–2.4 ms and $B_{\rm hexa} \simeq 0.8 \times 10^{16}$–$1 \times 10^{17}$ G. We note that a higher multipole order amplifies this systematic: since $B_l \propto f^{-l/2}$, the same factor of three in $f$ that moves a dipole field by 1.7 moves the hexapolar field by 5. The fitted pair ($P_0$, $B_{\rm hexa}$) should thus be read as one point on the family defined by Equation (11) rather than as two independently determined numbers. We also note that the 1 ms mass-shedding bound is itself EOS dependent and shifts with the assumed mass and radius (J. M. Lattimer & M. Prakash 2004, 2007; Y. Wang et al. 2024; R. Moradi et al. 2025; E. S. Yorgancioglu et al. 2025), and that $B_{\rm hexa}$ refers to the near-surface hexapolar component that governs the spin-down, not to the large-scale dipole or to the volume-averaged interior field discussed in Section 3.2.

A time-dependent term acts differently, and the resulting degeneracy cannot be removed by the X-ray light curve alone. Fallback accretion is the clearest example. The fallback rate declines roughly as $\dot{M} \propto t^{-5/3}$ (A. L. Piro & C. D. Ott 2011; R. Perna et al. 2014), so an accretion-powered luminosity would decay as $t^{-5/3}$, steeper than the observed $t^{-1.29}$; a pure fallback origin for the decay is thus not favored without additional time dependence. Fallback does not act only as a luminosity source, however. Accreted material exerts a torque on the star, and in the propeller regime it removes angular momentum at a rate set by $\dot{M}$ rather than by the magnetic braking law, which modifies the effective braking index and can mimic a different $l$ (B. P. Gompertz et al. 2013; S. L. Gibson et al. 2017, 2018). A slowly evolving radiative efficiency has the same effect: if $\eta_X \propto t^{-s}$, the measured index becomes $\alpha_X = 1 + 1/l + s$, and $l_{\rm eff}$ is biased low. The hexapolar fit is therefore one self-consistent realization of the decay, obtained under an explicit set of assumptions, rather than a determination of the magnetic field geometry of the engine, and the same data admit a fallback-regulated solution with a different $l$. The magnetar interpretation developed here consequently rests on the joint prompt, QPO, and afterglow phenomenology compared in Section 5, not on the afterglow fit in isolation.

The optical band carries the same message. R. Gupta et al. (2025) showed that the X-ray and optical afterglows of GRB 240825A cannot be fitted with a single standard synchrotron closure relation, and attributed the mismatch to radiative losses. A second reading is available: a common closure relation is not expected when the two bands are not powered in the same way, since an X-ray band carrying an internal, engine-powered contribution on top of the external-shock emission need not follow the same relation as an optical band dominated by the reverse and forward shocks (B. Zhang & P. Mészáros 2001; B. Zhang et al. 2006). C. Wu et al. (2025) reach a compatible conclusion from their own broadband modeling, finding that the standard afterglow model does not reproduce the X-ray light curve of GRB 240825A. In that case the fitted decay index encodes an energy-injection history, and Equation (8) constrains that history rather than isolating a magnetic multipole.

A further consistency check follows from the energy budget, in which the collimation correction enters both sides of the comparison. Requiring the rotational reservoir to cover the entire blast-wave energy is the strictest form of the test, since part of that energy may be supplied by the initial explosion rather than by spin-down. The reservoir is the rotational energy, $E_{\rm rot} = \frac{1}{2} I \Omega_{l,0}^2$, which for a newborn millisecond magnetar reaches $\sim$few $\times\ 10^{53}$ erg only in the maximally rotating, maximally massive case (B. D. Metzger et al. 2011; H.-J. Lü & B. Zhang 2014; M. G. Bernardini 2015). In the present model this reservoir is not an independent quantity: integrating Equation (4) gives $\int_0^\infty L_l\, dt = l\, L_{l,0} \tau_l = \frac{1}{2} I \Omega_{l,0}^2$, so the time-integrated isotropic X-ray energy is identically the rotational energy of the fitted star, $E_{X,\rm iso} = E_{\rm rot}(f = 1) \simeq 1.0 \times 10^{52}$ erg. The $f = 1$ fit therefore already assigns the whole reservoir to the observed isotropic X-ray output, and the corrected reservoir is $E_{\rm rot} = f E_{X,\rm iso}$.

Broadband modeling of the reverse- and forward-shock emission by C. Wu et al. (2025) gives an isotropic-equivalent kinetic energy $E_{\rm k,iso} = 5.25 \times 10^{54}$ erg. The outflow is collimated, so the energy the blast wave actually carries is

$$E_{\rm k} = E_{\rm k,iso}(1 - \cos\theta_{\rm j}) \simeq \frac{1}{2} E_{\rm k,iso}\, \theta_{\rm j}^2, \tag{12}$$

and the same factor $\theta_{\rm j}^2/2$ enters the reservoir through $f$, since $E_{\rm rot} = f E_{X,\rm iso}$. Forming the ratio,

$$\frac{E_{\rm k}}{E_{\rm rot}} = \frac{(\theta_{\rm j}^2/2)\, E_{\rm k,iso}}{(\theta_{\rm j}^2/2)(k/\eta_X)\, E_{X,\rm iso}} = \frac{\eta_X}{k} \frac{E_{\rm k,iso}}{E_{X,\rm iso}}, \tag{13}$$

the jet geometry cancels exactly, and the requirement $E_{\rm k} \leqslant E_{\rm rot}$ reduces to a condition on the radiative efficiency alone,

$$\eta_X \lesssim k \frac{E_{X,\rm iso}}{E_{\rm k,iso}} \simeq 2 \times 10^{-3}\, k. \tag{14}$$

Collimation lowers the energy demanded by the blast wave and the reservoir inferred from the fit by the same factor, so no assumption about the opening angle is needed: the budget constrains only the efficiency with which engine output is converted into the observed X-rays. Numerically, Equation (14) reads $\eta_X \lesssim 1\%$ for the bolometric correction $k \simeq 5$ that Y. Wang et al. (2024) estimate for the 0.3–10 keV band, and $\eta_X \lesssim 0.2\%$ if the fitted luminosity is already bolometric and $k = 1$. Both are comfortably inside the range explored by R. Moradi et al. (2025) and below the ceiling of D. Xiao & Z.-G. Dai (2019). The magnetar satisfies the energy budget, and it does so through the low radiative efficiency of the X-ray channel rather than through jet collimation.

The geometry is constrained as a by-product rather than as an input. The mass-shedding limit $P_0 \gtrsim 1$ ms caps the reservoir at $E_{\rm rot} \leqslant \frac{1}{2} I (2\pi/1\ {\rm ms})^2 \simeq 2.0 \times 10^{52}$ erg, and combining this with Equation (12) and $E_{\rm k} \leqslant E_{\rm rot}$ gives

$$\theta_{\rm j} \lesssim \left(\frac{2 E_{\rm rot}^{\rm max}}{E_{\rm k,iso}}\right)^{1/2} \simeq 5^\circ. \tag{15}$$

This is a prediction of the model rather than an assumption fed into it, and it is compatible with the lower limit on the jet opening angle obtained by C.-W. Wang et al. (2025) from the curvature effect in the prompt tail. The prompt phase is not restrictive either: applying the collimation correction to $E_{\rm iso} = 2.02 \times 10^{53}$ erg from Table 1 gives at most $7.7 \times 10^{50}$ erg at $\theta_{\rm j} = 5^\circ$. The energy budget therefore does not exclude a magnetar engine and it fixes the radiative efficiency the model must have through Equation (14).

### 3.2. Quasi-periodic Oscillation Consistency with a Magnetar Ce*ntral Engine

A magnetar central engine can naturally accommodate the $\sim$6.37 Hz QPO observed in GRB 240825A. A newborn millisecond magnetar possesses intrinsic geometric degrees of freedom capable of imprinting low-frequency modulation on a Poynting-dominated outflow (S. Xiao et al. 2024; G. P. Lamb et al. 2025).

For example, the free precession of a magnetically deformed neutron star can imprint periodic modulation on the outflow, with characteristic frequency $\nu_{\rm prec} \simeq \epsilon\, \nu_{\rm spin}$, where $\epsilon \equiv \Delta I / I_0$ is the stellar ellipticity, that is, the fractional difference between the principal moments of inertia about the symmetry axis and about an axis perpendicular to it. For a purely magnetic distortion $\epsilon \propto B^2$, so that $\epsilon$ quantifies the magnetic distortion (C. Cutler 2002; H. Gao et al. 2017; L. Xie et al. 2022). For a millisecond spin period, the observed $\nu_{\rm prec} \approx 6.37$ Hz requires $\epsilon \sim 10^{-2}$, a value that has been discussed in the context of magnetar precession candidates (K. Makishima et al. 2014; L. Zou & E.-W. Liang 2022; L. Zou & J.-G. Cheng 2024; G. Desvignes et al. 2024). We examine below whether such an ellipticity is physically attainable.

Crucially, free precession in a GRB central engine is not expected to remain coherent for a long time. In a newborn, rapidly rotating magnetar, strong coupling between the stellar interior and the magnetosphere, together with efficient bulk-viscous dissipation during the hot protoneutron star phase, can rapidly dissipate precessional energy and drive evolution of the inclination angle on timescales as short as seconds (S. K. Lander & D. I. Jones 2018; G. P. Lamb et al. 2025). As a result, precession is expected to briefly modulate the jet orientation and the Doppler factor of the photospheric emission over a narrow temporal window. This naturally accounts for both the short time of the observed QPO and its confinement to the thermal component of the prompt emission.

A widely studied framework of such geometric modulation invokes free precession in a magnetar whose electromagnetic output is dominated by spin-down losses. The afterglow fit of Section 3.1 favors a hexapolar configuration for the secular spin-down, and the dipolar form written below is used only as a convenient normalization over the QPO window. Across that window, the secular term changes the luminosity by a fraction $\simeq(1 + 1/l)\,\Delta t/\tau_l$, which is small compared with the observed modulation for any $l$ and for the fitted $\tau_l$, and which is in any case a smooth trend rather than a periodic signal. The secular braking law therefore enters Equation (16) as an effectively constant prefactor and drops out of the modulation analysis, whose frequency and amplitude are set by $\lambda$ alone. With that understood, the engine luminosity can be written as*

$$L_{\rm EM}(t) = \eta_X \frac{B_p^2 R^6 \Omega_0^4}{6c^3}\left(1 + \frac{t}{\tau_{\rm dip}}\right)^{-2} \lambda(\delta,\, \phi_0,\, k,\, \Omega_p), \tag{16}$$

where $\eta_X \gtrsim 0.1$ is the radiative efficiency (P. D. Lasky et al. 2017; H.-J. Lü et al. 2019), $\tau_{\rm dip}$ is the characteristic spin-down timescale, and $\lambda$ encodes geometric modulation induced by precession. Following A. Philippov et al. (2014), L. Arzamasskiy et al. (2015), and L. Zou & E.-W. Liang (2022), this factor can be expressed as

$$\begin{aligned}\lambda(\delta,\, \phi_0,\, k,\, \Omega_p) &= 1 + \delta \sin^2\phi \\ &\approx 1 + \delta[1 - (\cos\phi_0 + k(\cos\Omega_p t - 1))^2],\end{aligned} \tag{17}$$

where $\phi$ is the instantaneous inclination angle between the magnetic and rotation axes, $\phi_0$ is its mean value, $k \sim \mathcal{O}(1)$ depends on the Euler angles, and $\delta$ parametrizes the magnetospheric response. The precession frequency is related to the ellipticity through

$$\Omega_p \simeq \epsilon\, \Omega_{\rm spin} \cos\chi, \tag{18}$$

where $\chi$ is the inclination angle entering the free-precession relation (S. K. Lander & D. I. Jones 2017, 2018). Over the short $\sim$1 s QPO interval as in GRB 240825A, both $\Omega_{\rm spin}$ and $\Omega_p$ can be treated as effectively constant, so the detailed multipolar spin-down law does not materially affect the inferred precession frequency.

We apply Equation (16) to the 2.07–3.25 s interval during which the QPO is observed. The inclination angle may be written in the small-wobble limit as

$$\phi(t) = \phi_0 + \Delta\phi\, \sin(\Omega_p t + \varphi). \tag{19}$$

In this regime, the geometric factor $1 + \delta\sin^2\phi(t)$ produces a quasi-periodic modulation with peak fractional amplitude

$$A_\lambda \simeq \delta\, \Delta\phi\, \sin(2\phi_0). \tag{20}$$

For $\phi_0 \sim 40°$–$50°$, $\sin(2\phi_0)$ is close to unity, implying that the observed few-to-ten percent fractional rms amplitude requires $\frac{\delta\,\Delta\phi}{\sqrt{2}} \sim 0.03$–$0.1$. This can be achieved with modest wobble angles $\Delta\phi \sim 0.2$–$0.3$ rad together with $\delta \sim 0.3$–$0.5$, without requiring extreme geometric distortions. Since $A_\lambda$ is a ratio, $\eta_X$ and the entire prefactor of Equation (16) cancel from it, and the efficiency assumed there does not enter the inferred precession frequency or wobble angle; it is carried only so that Equation (16) returns an absolute luminosity. The symbol denotes two distinct conversions in this paper. In Equation (16) it is the prompt-phase conversion into the 100–300 keV band, for which values $\gtrsim$0.1 are standard, whereas in Equation (9) it is the conversion of spin-down power into the late-time 0.3–10 keV afterglow, a far less efficient channel that is typically one to two orders of magnitude smaller (D. Xiao & Z.-G. Dai 2019; R. Moradi et al. 2025). Only the latter propagates into the derived engine parameters, through the normalization.

The observed QPO frequency further implies, under the free-precession interpretation,

$$\epsilon \simeq \frac{\nu_{\rm prec}}{\nu_{\rm spin}\cos\chi} \simeq \frac{8.28\times10^{-3}}{\cos\chi}\left(\frac{\nu_{\rm prec}}{6.37\ {\rm Hz}}\right)\left(\frac{P_0}{1.38\ {\rm ms}}\right), \tag{21}$$

corresponding to $\epsilon \sim (1.1$–$1.4) \times 10^{-2}$ for $\chi \simeq 40°$–$50°$. However, standard magnetic-deformation estimates for a nonsuperconducting newborn magnetar (C. Cutler 2002; S. K. Lander & D. I. Jones 2018) give

$$|\epsilon_B| \approx 1.6\times10^{-6}\left(\frac{\langle B_t\rangle}{10^{15}\ {\rm G}}\right)^2, \tag{22}$$

where $\langle B_t\rangle$ is the rms volume-averaged internal toroidal field (C. Cutler 2002). Reaching $\epsilon \sim 10^{-2}$ would therefore require $\langle B_t\rangle \gtrsim 8\times10^{16}$ G, far exceeding the field strengths usually considered in realistic newborn magnetar models. The convective dynamo operating during the first seconds after core bounce saturates near equipartition with the convective kinetic energy, which sets an upper bound of order $3\times10^{17}$ G on the interior field (C. Thompson & R. C. Duncan 1993); three-dimensional crustal and core field-evolution calculations, together with population studies, place the fields actually realized in newborn magnetars well below this ceiling, at $\sim10^{15}$–$10^{16}$ G (K. N. Gourgouliatos et al. 2016; Q.-h. Peng et al. 2016; Y. Wang et al. 2024). Therefore, the free-precession interpretation from a magnetically deformed magnetar faces a significant energetic challenge and cannot be regarded as the default explanation for the observed QPO.

Notably, the above ellipticity estimations assume that the stellar ellipticity is produced entirely by magnetic deformation. In a newborn, rapidly rotating protoneutron star, this is unlikely to be the only contribution. Centrifugal flattening, differential rotation, anisotropic fallback accretion, starquake, and transient hydrodynamic asymmetries can all deform the star during the early engine phase (A. Colaiuda et al. 2008; L. Xie et al. 2022; E. Giliberti & G. Cambiotti 2022; Y. Gao et al. 2023). Rotational oblateness alone does not drive free precession in a perfectly axisymmetric configuration, but departures from axisymmetry or a misalignment between the spin and principal axes can in principle sustain transient precessional motion. This point is particularly relevant for GRB central engines, where the newborn neutron star may initially sit close to the bifurcation between the Maclaurin spheroid and Jacobi ellipsoid sequences. J. A. Rueda et al. (2022) showed that the $\nu$NS powering the early emission and afterglow of GRB 180720B and GRB 190114C may evolve from a short-lived triaxial Jacobi-like configuration into an axisymmetric Maclaurin spheroid through gravitational-wave emission, with an inferred initial eccentricity close to the bifurcation value of the Maclaurin sequence, $e \simeq 0.813$, corresponding to a highly oblate and a rapidly rotating configuration with millisecond or submillisecond rotation periods. More generally, deformed-magnetar models of GRB central engines allow ellipticities spanning $\epsilon \sim 10^{-5}$–$10^{-2}$, with the electromagnetic and gravitational-wave output depending sensitively on the stellar deformation, magnetic field geometry, and equation of state (P. D. Lasky & K. Glampedakis 2016; P. Hashemi et al. 2025). The effective ellipticity entering the free-precession estimate should therefore not be interpreted as a purely magnetic quantity. It is better regarded as an effective early-time nonaxisymmetry parameter that may encode contributions from magnetic stresses, rotationally induced deformation, fallback-driven distortions, and transient hydrodynamic asymmetries. A further difficulty for free precession is the confinement of the QPO to the photospheric component, and it is the same difficulty that the LT picture faces in Section 4. If the modulation arises from a geometric reorientation of the outflow, precession changes the beaming direction and the Doppler factor of the jet as a whole, and a 6.37 Hz signal should then appear in the contemporaneous nonthermal emission as well. Confining it to the thermal component requires the modulation to stay strong in the innermost emission zone and to be diluted or erased before the larger-radius dissipation regions. This can be arranged if the nonthermal region is extended enough to average over many cycles, but it is an additional requirement, and it applies to the magnetar free-precession picture as much as to the black hole one.

A physically plausible mechanism within the magnetar scenario is provided by global Alfvénic or magnetoelastic oscillations of the protoneutron star interior (Y. Levin 2006; H. Sotani et al. 2008; P. Cerdá-Durán et al. 2009). The characteristic magnetic frequency associated with such perturbations is set by the Alfvén crossing time across the stellar interior,

$$\nu_A \sim \frac{v_A}{2\pi R} = \frac{\langle B \rangle}{2\pi R\sqrt{4\pi\bar{\rho}}} \simeq 3.7\ \mathrm{Hz}\left(\frac{\langle B \rangle}{10^{15}\ \mathrm{G}}\right)\left(\frac{10^{14}\ \mathrm{g\ cm^{-3}}}{\bar{\rho}}\right)^{1/2}\left(\frac{12\ \mathrm{km}}{R}\right), \tag{23}$$

where $\bar{\rho}$ is the mean interior density. Inserting the hexapolar surface field $B_{\rm hexa} \simeq 2 \times 10^{16}$ G inferred from the afterglow modeling in Section 3.1 directly into Equation (23) yields $\nu_A \sim 75$ Hz, significantly above the observed 6.37 Hz. This naive estimate, however, applies to the surface multipolar component that dominates the spin-down energetics, rather than to the volume-averaged interior field associated with large-scale magnetic perturbations. A volume-averaged field of $\langle B \rangle \sim 1.5$–$2 \times 10^{15}$ G, consistent with a dipolar interior component coexisting with the stronger surface multipolar structure, yields magnetic crossing timescales in the observed QPO range directly from Equation (23) without requiring any additional tuning. Such interior field strengths are well within the range expected for newborn magnetars generated via a convective dynamo (C. Thompson & R. C. Duncan 1993) and are broadly compatible with the field geometry inferred for GRB 240825A.

We note that the Alfvén QPO mechanism, while successful in accounting for the frequencies of QPOs observed in mature Galactic magnetars (H. Sotani et al. 2008; P. Cerdá-Durán et al. 2009), has not been simulated for the extreme conditions of a protoneutron star. In a fluid core threaded by a magnetic field, the Alfvén continuum can cause rapid phase mixing and damping of global modes (Y. Levin 2006, 2007), and whether long-lived turning points or edge modes survive in a hot, crustless protoneutron star remains an open theoretical question. Moreover, for the inferred millisecond spin period, Coriolis forces and centrifugal deformation may substantially modify the mode spectrum, so that the relevant perturbations are more likely to belong to a mixed magnetoinertial spectrum rather than to purely Alfvénic modes (S. K. Lander et al. 2010). The crossing-time estimate used here should therefore be regarded as an order-of-magnitude guide rather than a precise eigenfrequency calculation; dedicated MHD simulations in the rapidly rotating protoneutron star regime would be required for a definitive prediction.

In this picture, the QPO modulation of the photospheric emission arises because magnetic perturbations propagating along the magnetospheric field lines transmit the interior oscillation pattern to the jet base, periodically modulating the photospheric luminosity and Doppler factor. The confinement of the modulation to the thermal component is naturally consistent with this interpretation. The photosphere, being located closest to the central engine, is expected to respond more directly to interior oscillation modes, whereas the nonthermal emission region at larger radii averages over many oscillation cycles and therefore suppresses coherent modulation.

The free-precession and magnetoelastic mechanisms are not mutually exclusive. Both can operate simultaneously in a newborn magnetar. However, on energetic and structural grounds, magnetic or magnetoinertial perturbations of the proto-neutron-star interior provide a less demanding explanation for the ∼6.37 Hz modulation observed in the thermal component of GRB 240825A than the free-precession interpretation, requiring only interior field strengths broadly compatible with the afterglow modeling rather than an ellipticity exceeding standard magnetic-deformation estimates by more than an order of magnitude. Outside the QPO interval, rapid damping of these perturbations (Y. Levin 2006; H. Sotani et al. 2008) and the transition toward a cleaner, dissipation-dominated outflow naturally suppress further observable modulation, consistent with the absence of any QPO signal beyond the 2.07–3.25 s interval.

## 4. Black Hole Interpretation

A rapidly spinning stellar-mass black hole, formed in either a collapsar or a compact-object merger, can power a relativistic jet via the BZ mechanism (R. D. Blandford & R. L. Znajek 1977), whereby rotational energy is extracted electromagnetically through large-scale magnetic fields threading the event horizon. The corresponding jet power is

$$L_{\rm BZ} \simeq 10^{50}\, a_*^2 \left(\frac{M_{\rm BH}}{3\,M_\odot}\right)^2 \left(\frac{B}{10^{15}\,{\rm G}}\right)^2 {\rm erg\ s^{-1}}, \tag{24}$$

where $a_*$ is the dimensionless spin parameter and $B$ is the characteristic magnetic field strength at the horizon.

Hyperaccretion rates of $\dot{M} \sim 0.01$–$0.1\,M_\odot\,{\rm s}^{-1}$can efficiently advect large-scale magnetic flux toward the central black hole. If sufficient magnetic flux accumulates, the inner accretion flow may enter a magnetically arrested disk (MAD) state, where the magnetic pressure becomes strong enough to regulate the inflow and enhance the extraction of rotational energy from a moderately spinning Kerr black hole (R. Narayan et al. 2003; A. Tchekhovskoy et al. 2011). In this regime, the BZ mechanism can, in principle, account for the prompt luminosity of GRB 240825A.

Among the proposed black hole–disk variability mechanisms capable of producing low-frequency QPOs, including pressure and gravity modes, orbital resonances, and magnetic instabilities, LT precession of a tilted inner accretion flow is the most quantitatively developed and the only one with a natural Hz-regime normalization for stellar-mass black holes at moderate radii; (see e.g., A. Ingram et al. 2009, and references therein). Other disk-origin mechanisms scale with the local Keplerian frequency,

$$\nu_{\rm K}(r) \simeq 1.08 \times 10^4\ {\rm Hz} \left(\frac{M_{\rm BH}}{3\,M_\odot}\right)^{-1} \left(\frac{r}{r_g}\right)^{-3/2}, \tag{25}$$

which places their characteristic variability in the $\sim$100–350 Hz range for inner-disk radii $r \sim 10$–$20\,r_g$. Producing a modulation at $\nu \sim 6.37$ Hz through direct Keplerian or local disk-orbital frequencies would require characteristic radii of $r \gtrsim 145\,r_g$, substantially larger than the compact inner-flow region usually associated with GRB central-engine activity. We therefore focus on LT precession as the most plausible black hole-based mechanism for the observed QPO.

In the weak-field limit, the LT precession frequency scales as

$$\nu_{\rm LT}(r) \simeq \frac{1}{2\pi}\frac{2GJ}{c^2 r^3} = \frac{a_*}{\pi}\frac{c^3}{GM_{\rm BH}}\left(\frac{r}{r_g}\right)^{-3}, \tag{26}$$

where $r_g \equiv GM_{\rm BH}/c^2$. For $M_{\rm BH} \sim 3\,M_\odot$ and $a_* \sim 0.6$, matching $\nu_{\rm QPO} \simeq 6.37$ Hz requires a characteristic precession radius of $r \sim 12$–$13\,r_g$ in this simplified scaling. More realistic rigid-body precession models, which integrate over the radial extent and surface-density profile of the precessing flow, yield comparable frequencies (A. Ingram et al. 2009; A. Ingram & C. Done 2011), and GRMHD simulations confirm that a sufficiently tilted disk can drive a quasi-periodically precessing jet (M. Liska et al. 2018).

A central difficulty, however, is that the LT modulation is generated in the innermost accretion flow and must propagate coherently through a relativistic jet over many orders of magnitude in radius before producing observable emission. In GRB 240825A, the photospheric emission is expected to emerge at radii of order $r_{\rm ph} \sim 10^{11}$–$10^{13}$ cm, while nonthermal dissipation may occur farther out still. Any periodic signal imprinted at $r \sim 12\,r_g \sim 5 \times 10^6$ cm must therefore survive propagation over roughly five to seven orders of magnitude in radius. Extended or multizone dissipation regions can naturally wash out coherent modulation through phase mixing, angular averaging, and differential photon travel times across the emitting surface. Whether the precession signal survives this propagation depends sensitively on the jet structure, angular emissivity profile, and Lorentz-factor evolution, none of which are tightly constrained in GRB 240825A.

Moreover, in GRB 240825A, the 6.37 Hz modulation is detected only in the 100–300 keV band, where the quasi-thermal photospheric emission dominates. Although variability at lower-frequency ($\sim$0.67 Hz) has also been reported in the nonthermal component (G.-Y. Li et al. 2025), no corresponding QPO 6.37 Hz is measured in the nonthermal emission. However, a purely geometric LT-precession mechanism would be expected to affect multiple emission regions (B.-Q. Huang & T. Liu 2022), making the observed frequency dependence difficult to explain within the simplest precession picture, although it does not formally exclude it.

An additional tension arises from the frequency stability of the signal. Since $\nu_{\rm LT} \propto r^{-3}$, a modest inward migration of the inner disk by a factor of $\sim$1.15 in radius over the $\sim$1 s QPO interval would shift the precession frequency by $\sim$50%, well above the observed stability of the 6.37 Hz signal across multiple cycles. Maintaining a stable precession frequency therefore requires that the characteristic radius of the inner flow remain nearly constant during the QPO interval, which in turn places stringent constraints on the accretion-rate evolution and disk structure during the prompt phase.

Finally, while the X-ray afterglow does not by itself discriminate between engine models, a black hole interpretation requires fallback- regulated accretion or time-dependent jet efficiency to reproduce the subsequent decay. This introduces a second independent mechanism, operating on timescales and radii very different from those responsible for the prompt QPO. The degree to which these two components can be unified within a single black hole-accretion framework is a key consideration in the model comparison presented in Section 5.

## 5. Comparison of Central Engine Models

Both magnetar and black hole central engines can, in principle, produce a hertz-level QPO and sustain energy injection into the afterglow of GRB 240825A. The key question is whether a single physical framework can simultaneously account for the evolution of the well-separated spectral components, the strongly magnetized ejecta inferred from the reverse-shock analysis, the afterglow decay properties, the low-frequency QPO, and its confinement to the photospheric component without invoking separate and independent mechanisms for each feature.

In the LT-precession scenario, producing a strong ∼6.37 Hz modulation requires that the inner accretion flow remain geometrically thick, approximately rigid, and only weakly affected by alignment torques over several precession cycles. Whether these conditions persist in the rapidly evolving, highly magnetized environment of a GRB prompt phase is uncertain. As discussed in Section 4, the spectral selectivity of the QPO, confined to the photospheric component and absent in the nonthermal emission, is difficult to explain through a purely geometric jet reorientation mechanism, which would generally imprint modulation across multiple spectral components rather than remaining confined primarily to the photospheric emission. This difficulty is not specific to the black hole. Magnetar free precession is also a geometric reorientation mechanism and inherits it, as discussed in Section 3.2, which is one reason to prefer the magnetoelastic channel within the magnetar scenario, where the modulation is imposed at the base of the outflow rather than on its orientation. An additional tension arises from the frequency stability of the signal: since $\nu_{\rm LT} \propto r^{-3}$, even modest inward migration of the inner disk during the prompt phase would produce a measurable frequency drift that is not observed. Fallback-regulated accretion or time-dependent jet efficiency are then required to explain the afterglow, introducing a second independent mechanism.

By contrast, a magnetar can unify both phases into a single framework. The hexapolar spin-down model reproduces the afterglow decay quantitatively, with physically reasonable initial parameters ($P_0 \simeq 1.38$ ms, $B_{\rm hexa} \simeq 2.0 \times 10^{16}$ G), understood as one point of the one-parameter family of Equation (11) rather than as independently determined values. The QPO stems from magnetic or magnetoinertial perturbations of the protoneutron star interior, which naturally introduce magnetic timescales in the observed Hz-regime for volume-averaged interior fields $\langle B \rangle \sim 1.5\text{–}2 \times 10^{15}$ G, consistent with convective-dynamo expectations and broadly compatible with the afterglow modeling. The confinement of the QPO to the photospheric component follows naturally, since the photosphere is the emission zone most directly coupled to the engine. The large magnetic field ratio $R_B \sim 300$ inferred from the reverse-to-forward shock transition further supports a strongly magnetized ejecta (C. Wu et al. 2025), consistent with a magnetar-driven outflow.

As shown in Section 3.1, the afterglow fit does not on its own identify the engine. The decay index constrains an effective braking index, which can also be produced by alternative scenarios, such as a fallback-regulated flow with a propeller torque or a time-dependent radiative efficiency (B. P. Gompertz et al. 2013; S. L. Gibson et al. 2017, 2018). The hexapolar solution is one realization of the decay rather than a measurement of the field structure, and the comparison made here therefore rests on the prompt and QPO phenomenology as much as on the light curve. In this context, fallback regulation is an alternative to multipolar spin-down in the magnetar case and a requirement in the black hole case, where no intrinsic spin-down clock reproduces the decay. What distinguishes the two models is therefore which ingredients each must add, not whether fallback occurs.

We note that it is unlikely for the X-ray afterglow alone to discriminate between the two engines. Recent systematic studies have shown that rapidly rotating black holes through spin-down models can extract energy via MADs and reproduce similar decay correlations, including plateaus, with statistical significance comparable to that of magnetar-based models (A. Ł. Lenart et al. 2025). In the context of GRB 240825A, discriminating between the engines only becomes possible once considering the prompt QPO and the afterglow together. In the magnetar framework, these observational features are governed by the same long-lived compact object, within a common magnetized-engine framework. In the black hole framework, they require separate mechanisms operating at very different radii and timescales.

Taken together, the combined prompt and afterglow phenomenology of GRB 240825A is most self-consistently explained by a magnetar central engine. A black hole origin cannot be excluded, particularly for the prompt QPO considered in isolation, but it is comparatively less parsimonious.

## 6. Conclusions

We have examined and compared the central-engine physics of GRB 240825A within magnetar and black hole frameworks using its multicomponent prompt spectrum, X-ray afterglow, and the $6.37 \pm 0.05$ Hz QPO detected in the photospheric-dominated band during the 2.07–3.25 s interval. Our main findings are as follows;

The X-ray afterglow is quantitatively reproduced by a multipolar magnetar spin-down model dominated by a hexapolar component, yielding an initial spin period $P_0 \simeq 1.38$ ms and a surface magnetic field $B_{\rm hexa} \simeq 2.04 \times 10^{16}$ G. Adding a dipole component does not significantly improve the fit because the local slope of the spin-down law never exceeds its asymptotic value, the measured decay index bounds the dominant order from above at $l \leqslant 1/(\alpha_X - 1) = 3.4 \pm 0.2$, which excludes an octupolar wind outright, while the absence of curvature across the fitted range excludes the dipolar and quadrupolar cases, whose slopes would still have to steepen by 55% and 16%, respectively. What this constrains is a braking index, and it does not by itself separate a multipolar wind from a fallback-regulated flow. The quoted $P_0$ and $B_{\rm hexa}$ correspond to isotropic emission with unit efficiency; a beaming, efficiency, and bolometric rescaling $f$ moves them along $P_0 \propto f^{-1/2}$ and $B_{\rm hexa} \propto f^{-3/2}$. Beaming and radiative efficiency enter $f$ with opposite exponents and largely cancel, leaving $f \simeq 1$ with a residual factor of about three, which corresponds to $P_0 \simeq 1.0\text{–}2.4$ ms and $B_{\rm hexa} \simeq 0.8 \times 10^{16}\text{–}1 \times 10^{17}$ G. In the afterglow energy budget, the collimation correction cancels identically between the blast wave and the inferred reservoir, so that the budget constrains only the radiative efficiency, $\eta_X \lesssim 2 \times 10^{-3} k$, that is 0.2%–1% for $k = 1\text{–}5$. Combined with the mass-shedding limit, the same budget bounds the jet half-opening angle at $\theta_{\rm j} \lesssim 5^\circ$ as a derived consequence.

Within the magnetar scenario, magnetic or magnetoinertial perturbations of the protoneutron star interior can plausibly account for the 6.37 Hz QPO. For volume-averaged interior fields $\langle B \rangle \sim 1.5\text{–}2 \times 10^{15}$ G, the Alfvén crossing-time estimate introduces characteristic magnetic timescales in the observed frequency range without requiring extreme stellar deformations. By contrast, free precession through magnetic deformation demands an ellipticity $\epsilon \sim 10^{-2}$ that exceeds standard magnetic-deformation estimates by more than an order of magnitude. However, magnetic deformation is unlikely to be the only origin of the ellipticity in free-precession mechanisms.

The confinement of the 6.37 Hz QPO to the quasi-thermal photospheric component, while absent in the nonthermal emission, indicates that the modulation originates close to the central engine and is preserved most efficiently in the innermost emission zone. This spectral selectivity is consistent with an engine-coupled magnetic or magnetoinertial perturbation picture, and disfavors a purely geometric origin, such as LT jet precession and magnetar free precession, which would generally be expected to imprint modulation across multiple spectral components rather than a single component.

A Kerr black hole driving a BZ jet remains a viable alternative. LT precession of a tilted inner flow can produce Hz-level QPOs for plausible black hole masses and spins, and the afterglow can be reproduced with fallback-regulated accretion or time-dependent jet efficiency. However, these require two independent mechanisms for features that the magnetar accounts for under a single, unified framework. The distinction is that fallback regulation is optional in the magnetar case and necessary in the black hole case, not that fallback is absent in one of them. The frequency stability of the observed QPO and its spectral confinement requires additional assumptions that are not naturally satisfied in the LT-precession scenario.

In summary, the combined prompt and afterglow properties of GRB 240825A are most self-consistently explained by a long-lived magnetar central engine. A definitive discrimination between the two scenarios would require either a direct gravitational-wave detection of the characteristic spin-down signal or a demonstrated correlation between the QPO phase and the polarization angle of the photospheric emission, measurements that next-generation facilities such as POLAR-2, Einstein Telescope, and Cosmic Explorer are positioned to provide.

## Acknowledgments

R.M. acknowledges support from the Institute of High Energy Physics, Chinese Academy of Sciences (E32984U810) and the Beijing Natural Science Foundation (IS24021). This work made use of data supplied by the UK Swift Science Data Centre at the University of Leicester and from the Fermi Gamma-ray Burst Monitor (GBM), provided by the NASA/Goddard Space Flight Center. The authors acknowledge the use of AI for necessary language refinements. Following this assistance, the authors manually cross-checked the originality and accuracy of the content, and they maintain full accountability for the final manuscript.

## Appendix A
## Bolometric Fluence

The bolometric fluence, $S_{\rm bolo}$, is defined as the total energy per unit area integrated over $T_{90}$ duration and energy range, which is expressed as

$$S_{\rm bolo} = \int_{E_{\rm min/1+z}}^{E_{\rm max/1+z}} E\frac{dN}{dE}dE. \tag{A1}$$

Here, the $E_{\rm min}$ and $E_{\rm max}$ range from 8 keV to 30 MeV, respectively. From the $k$-corrected bolometric fluence $S_{\rm bolo}$, we can then proceed to calculate the Isotropic-equivalent energy, expressed as

$$E_{\rm iso} = \frac{4\pi d_L^2 S_{\rm bolo}}{1+z}, \tag{A2}$$

where $d_L^2$ is the luminosity distance, as follows

$$d_{\rm L} = (1+z)\frac{c}{H_0}\int_0^z \frac{dz'}{\sqrt{\Omega_{\rm m}(1+z')^3 + \Omega_\Lambda}}. \tag{A3}$$

## Appendix B
## Cosmological Parameters

We calculated the luminosity distance using cosmological parameters from Planck Collaboration et al. (2020) adopting a flat ΛCDM cosmology with Hubble constant $H_0 = 67.4\ {\rm km\ s^{-1}\ Mpc^{-1}}$, matter density parameter $\Omega_{\rm m} = 0.315$, and dark-energy density $\Omega_\Lambda = 1 - \Omega_{\rm m} = 0.685$.

## Appendix C
## Spectral Models

### *C.1. The Band Function*

The GRB spectrum can be fit with a smoothly joint broken power law. The photon number in this model is expressed as

$$N(E) = \begin{cases} A\left(\frac{E}{100}\right)^{\alpha}\exp\left(-\frac{E}{E_0}\right), & E < (\alpha-\beta)E_0 \\ A\left[\frac{(\alpha-\beta)E_0}{100}\right]\exp\left(\frac{E}{100}\right)^{\beta}, & E \geqslant (\alpha-\beta)E_0 \end{cases} \tag{C1}$$

where ($A$) is the normalization factor, ($\alpha$) is a low-energy spectral index, ($\beta$) is the high-energy spectral index, and ($E_0$) is the break energy. The peak energy in the $E^2N(E)$ spectrum is called the $E_p$, which is related to $E_0$ through $E_p = (2+\alpha)E_0$ (B. Zhang 2018).

### *C.2. Cutoff Power Law*

For the first portion of the Band function, the photon number spectrum reads as

$$N(E) = A\left(\frac{E}{100}\right)^{\alpha}\exp\left(-\frac{E}{E_c}\right), \tag{C2}$$

where $A$ is the normalization factor, $\alpha$ is the low-energy spectral index, $E_c$ is the characteristic energy and the $E_p$ is defined similar to the Band function $E_p = (2+\alpha)E_c$.

### *C.3. Power Law*

This is one of the simplest models in spectral analysis and the photon number spectrum is described as

$$N(E) = A\left(\frac{E}{100}\right)^{\alpha}, \tag{C3}$$

where $A$ is the normalization factor and $\alpha$ is the spectral index.

## ORCID iDs

Daban Mohammed Saeed https://orcid.org/0009-0009-1933-3204
Rahim Moradi https://orcid.org/0000-0002-2516-5894
Chen-Wei Wang https://orcid.org/0009-0008-8053-2985